\documentclass[12pt]{article}
\usepackage{graphicx}
\begin{document}

\begin{center}
{\large \bf Electrical Conductivity in General Relativity} \\
\vspace*{9 mm} {\bf B.J. Ahmedov$^{1,2,3}$ and M.J.
Ermamatov$^{1,3}$}\\

\vspace*{9 mm}

{\it $^{1}$Institute of Nuclear Physics and Ulugh Beg Astronomical
Institute\\ Astronomicheskaya 33,
    Tashkent 700052, Uzbekistan\\}
    {\it $^{2}$International Center for
Relativistic Astrophysics, Pescara, Italy}\\ {\it $^{3}$
Inter-University Centre for Astronomy and Astrophysics\\
         Post Bag 4 Ganeshkhind,
411007 Pune, India}
\end{center}

\vspace*{20mm}

{\small The general relativistic kinetic theory including the effect of a
stationary gravitational field is applied to the electromagnetic
transport processes in conductors. Then it is applied to derive the
general relativistic Ohm's law where the gravitomagnetic terms are
incorporated. The total electric charge quantity
and  charge  distribution
inside conductors carrying  conduction  current  in some relativistic
cases are considered.
The general relativistic Ohm's law is applied to predict new
gravitomagnetic and gyroscopic effects which can, in principle, be used to
detect the Lense-Thirring and rotational fields.}
\newpage
\section{INTRODUCTION}

Kinetic theory in general relativity is treated by a number of authors, and
has been reviewed in much more detail by Ehlers$^{(1)}$, using the
differential calculus on Riemannian manifolds. Recently the growing interest
to the general relativistic kinetic theory in plasma appeared$^{(2)-(4)}$.
However, the kinetic theory of electrical conductivity in conductors
in the presence of gravitational field is not
strongly discussed in the literature and
its presentation may be of some interest
(The formulation of the kinetic theory of electrical conductivity in the
astrophysical matter in the flat spacetime can be found, for example,
in$^{(5)}$).
In this paper we consider the general relativistic kinetic theory
needed to determine on microscopic level the influence of a stationary
gravitational field on electromagnetic transport phenomena for normal
conductors.
This approach is interesting, even from purely theoretical point of
view, because it involves the interplay of gravitation, electromagnetism,
relativity and statistical physics.

The paper is organized in the following way.
First, in $\S2$ we consider Boltzmann's
kinetic equation in a manifestly covariant
form and the distribution function for electron gas in the local
equilibrium. Then in the approximation of relaxation time we derive the
general-relativistic nonequilibrium distribution function for electron gas
in an applied electromagnetic field which is the key element in the
construction of microscopic theory of
conductivity: the general relativistic Ohm's law is defined by the
corresponding average in velocity space.
In $\S3$ it is shown that the theoretical project$^{(6)}$ on
measuring gyroscopic effect in two series connected coil
conductor carrying current
and rotating in opposite directions has to give the null result.
It appears author of the paper$^{(6)}$ ignored the surface charges which
compensate rotationally induced space charges.  Finally, we apply the derived
Ohm's law to a few devices that can be used as detectors of gravitomagnetism
and rotation.

\section{GENERAL RELATIVISTIC KINETIC THEORY OF ELECTRICAL CONDUCTIVITY}

General relativistic equations of macroscopic electrodynamics
can be derived in two steps.
First, a suitable averaging process must be performed so as to
obtain the Maxwell equations, being valid at the macroscopic level, from the
fundamental Maxwell-Lorentz equations found at the microscopic one

\begin{eqnarray}
\label{lorentz}
e^{\alpha\beta\mu\nu} f_{\beta\mu ;\nu}= 0,\hspace{10mm}
{f^{\alpha\beta}}_ {;\beta}= \frac{4\pi}{c}j^\alpha,
\end{eqnarray}
which are general covariant in their form, where
in the right-hand side - a microscopic four-current density
 $j^\alpha (x)$,
and in the left-hand side - the covariant derivative of microscopic tensor
of electromagnetic field $f_{\mu\nu}$. Here Greek indices go through
$0,1,2,3$, $e^{\alpha\beta\mu\nu}$
is the tensorial expression for the Levi-Civita symbol$^{(7)}$.

We shall neglect all the feed back of the electromagnetic field on the
metric
itself under the well justified assumption that the electromagnetic energy
usually can be considered negligible with respect to the mass energy of
source of gravitational field. Averaging out the Maxwell-Lorentz
equations in the Riemannian background gives

\begin{eqnarray}
e^{\alpha\beta\mu\nu} F_{\beta\mu ,\nu}= 0,\hspace{10mm}
{F^{\alpha\beta}}_ {;\beta}= \frac{4\pi}{c}<j^\alpha>,
\end{eqnarray}
where $F_{\mu\nu}\equiv <f_{\mu\nu}>$ is averaged macroscopic tensor of
electromagnetic field. Operations of averaging out
and covariant differentiation are assumed to
commutate (although it can not be
justified in the general case, see, for example,$^{(8)}$).

Denoting the particles by twice notation $ai$, where $a$ is
number of atom consisting of particles $i$, we can write the current
density $j^\alpha(x)$ as

\begin{eqnarray}
\label{microcurrent}
j^\alpha (x)=\sum_{a,i}^{} q_{ai} \int_{+ \infty}^{- \infty} u_{ai}^\alpha
(\sigma _{ai})\delta ^4(x-z_{ai}(\sigma _{ai}))d\sigma _{ai},
\end{eqnarray}
$q_{ai}$ is the charge of particle $ai$, $\sigma_{ai}$ its proper time, and
$u^\alpha _{ai}$ its four-velocity.

Choose now world line $z^\alpha _a$ with proper parameter $\sigma _a$
which
describes motion of atom as whole. The relative position of a particle in
atom

\begin{eqnarray}
\label{position}
s^\alpha _{ai}\equiv z^\alpha _{ai} - z^\alpha _a
\end{eqnarray}
is defined on the hypersurface $\sigma _a = const$: $s^\alpha _{ai}u_\alpha
=
0$.

Substituting expression (\ref{position}) into (\ref{microcurrent}) and
expanding $\delta$- function in
powers of $s^\alpha _{ai}$ one can obtain

\begin{eqnarray}
j^\alpha (x)=\sum_{a}^{}  \int_{+ \infty}^{- \infty} d\sigma _a \sum q_i
(u_a^\alpha + \frac{ds^\alpha _{ai}}{d\sigma _a})
\{\delta ^4(x-z_a(\sigma _a))- s^\rho _{ai}\nabla _\rho \delta ^4(x-z_a)\}\ ,
\end{eqnarray}
when only first terms of expansion are kept.
First term of expansion called the free current density has the form

\begin{eqnarray}
j^\alpha _f (x)=\sum_{a}^{} q_a \int_{+ \infty}^{- \infty} d\sigma _a
u_a^\alpha (\sigma _a) \delta ^4(x-z_a(\sigma _a)).
\end{eqnarray}
One can show that the last terms of expansion are the divergence of an
antisymmetric tensor of polarization and magnetization of atom $m_a^{\mu
\nu}$ (as the general relativistic generalization of Kauffmann's
method$^{(9)}$)

\begin{eqnarray}
j^\alpha _{(in)} = \nabla _\beta \sum_{a} \int d\sigma _a m_a^{\alpha\beta}
\delta ^4(x-z_a),
\end{eqnarray}
where $m_a^{\mu \nu}=\pi_a^\nu \wedge u_a^\mu -\mu_a^{\mu\nu}$,
$\pi_a^\nu =\sum_{i} q_i s^\nu _{ai}$ is the electric dipole moment,
$\mu _{a}^{\mu \nu}= \frac{1}{2} \sum_{i} q_i s^\nu _{ai} \wedge
\frac{ds^\mu _{ai}}{d \sigma _a}$ is the covariant magnetic dipole moment,
$\wedge$ denotes the wedge product, $\nabla _\beta $ represents the
covariant
derivative.

Now we introduce the microscopic dipole density

\begin{eqnarray}
m^{\alpha\beta}(x)=\sum_{a} \int d\sigma _a m_a^{\alpha\beta}
\delta ^4(x-z_a),
\end{eqnarray}
so that current $<j^\alpha >$ becomes
\begin{eqnarray}
<j^\alpha >=<j_f^\alpha + \nabla _\alpha m^{\alpha\beta}>.
\end{eqnarray}

Upon averaging (and denoting macroscopic quantities by capital letters)
with
the conventional introduction of macroscopic tensor of electromagnetic
induction

$$H^{\alpha\beta}=F^{\alpha\beta}-\frac{4\pi}{c}M^{\alpha\beta}$$
one can  obtain Maxwell equations

\begin{eqnarray}
\label{maxwell}
e^{\alpha\beta\mu\nu}F_{\beta\mu,\nu}=0, \qquad
{H^{\mu\nu}}_{;\nu}=\frac{4\pi}{c}J^\mu
\end{eqnarray}
being valid for an arbitrary medium in general relativity.
Thus the standard form of Maxwell equations is indeed
valid in the Riemannian background, if suitable care is taken in the
definition
of macroscopic quantities, that is electric and magnetic moments are
incorporated into a general covariant scheme.

Furthermore the electric current $J^\alpha$ is still the sum of two terms
corresponding to the convection current and to the conduction
current $\hat j^\alpha$, respectively

\begin{eqnarray}
J^\alpha= c\rho_0 u^\alpha +\hat j^\alpha,\quad \hat j^\alpha
u_\alpha\equiv 0,
\end{eqnarray}
$\rho_0$ is the proper
density of free electric charges.

The system of differential equations (\ref{maxwell}) can be closed if and
only if the
general-relativistic constitutive relations between the
inductions and fields, on the one hand, conduction current and field
characteristics, on the another hand, through the
medium characteristics are given. Thus as a second step material couples
which depend on the nature of matter should be defined.

For the special case of
conducting media the conduction current is connected with the
electromagnetic field tensor $F^{\alpha\beta}$ according to the
generalized Ohm law which can be obtained with help of
the classical kinetic theory of a gas in the external gravitational
field (a special case of such a gas is the conduction electrons in a
metal.).

Kinetic theory can be constructed on the basis of invariant distribution
function $f(x,u)$ which depends on the coordinates $x^{\alpha}$ and
velocities
$u^{\alpha}$ and is described by the general-relativistic Boltzmann's
kinetic equation  for the electron gas$^{(1,10)}$

\begin{equation}
\frac{\partial f}{\partial x^\alpha}u^\alpha +
\frac{\partial f}{\partial u^\alpha}\frac{du^\alpha}{d\sigma}
=J(x,u).
\label{xal}
\end{equation}

In the general case the collision integral $J(x,u)$ has complicated form and
the kinetic equation (\ref{xal})
is the integro-differential one with respect
to the distribution function of electrons in the metal.
However in the case of the statistical local equilibrium the
collision integral vanishes since in that case the growing of
entropy is equal to zero$^{(11)}$
\begin{equation}
S^\mu(x)=-\int\frac{d^3u}{u^0}u^{\mu}f(x,u)[lnf(x,u)-1]\ ,
\end{equation}
that is when the distribution function satisfies to the
functional expression
\begin{equation}
f(x,u)f(x,u_1)=f(x,u')f(x,u_1')\ ,
\end{equation}
where $u, u_1$ and $u', u'_1$ are velocities of colliding particles
before and after the collision, respectively.

Consequently the equilibrium distribution function for the electron gas
subjects to the following kinetic equation
\begin{equation}
\frac{\partial f_0}{\partial x^\alpha}u^{\alpha}-
\Gamma^{\alpha}_{\beta\gamma}u^{\beta}u^{\gamma}
\frac{\partial f_0}{\partial u^\alpha}=0\ ,
\end{equation}
where $f_0$ is the equilibrium distribution function.

The local equilibrium of particles can take place only in
the stationary gravitational field when spacetime allows the
existence of the timelike Killing vector
\begin{equation}
\xi_\alpha=\Xi^{1/2}\lambda_\alpha;\qquad (\xi\xi)=-\Xi\ ,
\end{equation}
$\lambda_\alpha$ is the 4-velocity of the conductor as a whole.

It is known that the symmetry of space-time (the existence of the
Killing vector) corresponds to the conservation law.
In fact, the quantity
\begin{equation}
\frac{\delta(u\xi)}{\delta\tau}
=u^{\mu}_{;\nu}u^{\nu}\xi_{\mu}+\xi_{\mu;\nu}
u^{\mu}u^{\nu}=0
\end{equation}
is constant along the geodesic line and can be interpreted
as conserved energy of particle. Then the equilibrium distribution
function takes the form$^{(10)}$
\begin{equation}
f_0(x,u)=\exp\{\beta +(\xi(x)u)\}\ .
\end{equation}

Consider now the equilibrium distribution function of the
conduction electrons
\begin{equation}
f_0(x,u)=\exp\{\beta +\frac{e}{mc^2}(A(x)\xi)+(\xi u)\}
\end{equation}
in the presence
of electromagnetic field which satisfies the kinetic equation
\begin{equation}
\frac{\partial f_0}{\partial x^\alpha}u^{\alpha}-
\Gamma^{\alpha}_{\beta\gamma}u^{\beta}u^{\gamma}
\frac{\partial f_0}{\partial u^\alpha}
+\frac{e}{mc^2}F^{\alpha\sigma}u_{\sigma}
\frac{\partial f_0}{\partial u^\alpha}=0\ ,
\end{equation}
if the electromagnetic field is stationary i.e.
${\pounds}_\xi F_{\alpha\beta}=
0$ and a gauge is chosen in a way when ${\pounds}_\xi A_\alpha =0$,
where ${\pounds}_\xi$ is the Lie
derivative with respect to $\xi^\alpha$ and $A_\alpha$ is the vector
potential of electromagnetic field.

The normalization constant $\beta$ can be chosen in the following form
\begin{equation}
f_0(x,u)=\exp\left\{\frac{\tilde\mu(x)+(\xi(x)u)}{K\tilde T(x)}
\right\}\ ,
\end{equation}
where $\tilde\mu(x)=\Xi^{1/2}\mu(x)=\tilde\zeta +eA_\alpha\xi^\alpha$
is the related gravitoelectrochemical potential, $\zeta$ is the ordinary
chemical potential, $T(x)$ is the temperature measured by
observer whose 4-velocity field is $\lambda_\alpha$. The following condition
\begin{equation}
\label{equilibrium}
\tilde\mu_{,\alpha}-\frac{\tilde\mu-\epsilon}{\tilde T(x)}
\tilde T_{,\alpha}=
\zeta_{,\alpha}-w_\alpha\zeta+eE_\alpha-\frac{\tilde\mu-\epsilon}
{\tilde T(x)}\tilde T_{,\alpha}
=0 \ ,
\end{equation}
takes place during the thermodynamical equilibrium.
Here $\epsilon =-\xi u$ is the Fermi energy, $\Xi^{1/2}_{,\alpha}=
-\Xi^{1/2}\lambda_{\alpha;\beta}\lambda^\beta =-\Xi^{1/2}w_\alpha$,
$w_\alpha$ is the absolute acceleration of the conductor,
$E_\alpha$ is the electric field as measured by observer at rest
relative to the conductor. Equation (\ref{equilibrium}) indicates
the well-known fact that inner electric field can be induced inside
the conductor even in the absence of a current,
see, for example$^{(12)}$

Assume that nonequilibrium
distribution of velocities differs from the equilibrium one due to the
effect of an applied field and relaxate to the equilibrium state
exponentially with time $t$

\begin{equation}
\frac{\partial f}{\partial t}=J=\frac{\partial (f-f_0)}{\partial t}=
-\frac{f-f_0}{\tau (x,u)},
\label{fJ}
\end{equation}
where $\tau (x,u)$ is the relaxation time.
The solution of the equation
(\ref{fJ}) has form

\begin{equation}
(f-f_0)_t=(f-f_0)_{t=t_0}\exp(-t/\tau).
\end{equation}

Assume the deviation caused by external applied field is small, i.e.
\begin{equation}
f-f_0=f_1<<1.
\label{approx}
\end{equation}

According to (\ref{fJ}) one can now rewrite the kinetic equation
(\ref{xal}) in the form
\begin{equation}
\frac{\partial f}{\partial x^\alpha}u^{\alpha}-
\Gamma^{\alpha}_{\beta\gamma}u^{\beta}u^{\gamma}
\frac{\partial f}{\partial u^\alpha}
+\frac{e}{mc^2}F^{\alpha\sigma}u_{\sigma}
\frac{\partial f}{\partial u^\alpha}+\frac{f_1}{\tau}=0\ .
\end{equation}

In the linear approximaion in expansion of distrubution
function we found that
the nonequilibrium function for electron gas in a static
gravitational field and zero magnetic field is
\begin{equation}
f_1=\tau\frac{\partial f_0}{\partial \epsilon}
\frac{v^{\alpha}}{\sqrt{1-\frac{v^2}{c^2}}}\left\{
\frac{\partial \tilde{\mu}}{\partial x^{\alpha}}-
\left(
\frac{\tilde\mu(x)-
\epsilon}{\tilde{T}{(x)}}
\right)
\frac{\partial \tilde{T}{(x)}}{\partial x^\alpha}\right\}\ .
\label{vlao}
\end{equation}

Inserting (\ref{vlao}) into the expression for 4-current $j^{\alpha}$
one can get the following expression
$$
j^{\alpha}=\frac{2e}{\hbar^3}\int{v^\alpha f_1d^3p}=
$$
\begin{equation}
\frac{2e^2\tau}{\hbar^3}\left\{
\frac{1}{e}
\frac{\partial \tilde{\mu}}{\partial x^\alpha}-
\frac{1}{e}
\left(
\frac{\tilde{\mu}(x)-\epsilon}{\tilde{T}(x)}
\right)
\frac{\partial \tilde{T}}{\partial x^\rho}+F_{\rho\sigma}
\lambda^\sigma
\right\}\
\int \frac{v^{\alpha}v^\rho}{\sqrt{1-\frac{v^2}{c^2}}}
\frac{\partial f_0}{\partial \epsilon}d^3p.
\label{ja}
\end{equation}
The the conduction current is aligned along the direction of
the electric field. Taking into account this fact one can
write (\ref{ja}) as$^{(13)}$
\begin{equation}
j_{\alpha}=\sigma\left\{
\frac{\partial \tilde{\mu}_e}{\partial x^\alpha}-
\left(
\frac{\tilde{\mu}(x)-\epsilon}{\tilde{T}(x)}
\right)
\frac{\partial \tilde{T}}{\partial x^\alpha}+F_{\alpha\sigma}
\lambda^\sigma
\right\}\ ,
\end{equation}
where $\sigma=\frac{2e^2\tau}{\hbar^3}
\int{\frac{v^2}{\sqrt{1-\frac{v^2}{c^2}}}
\frac{\partial f_0}{\partial \epsilon}d^3p}$ is
a coefficient of electrical conductivity.

Let us estimate the coefficient of electrical
conductivity in the approximation $\frac{v}{c}<<1$. In this case,
$p=\sqrt{2\epsilon m}$, $d^3p=4\pi p^2dp=2\pi m^3\sqrt{2\epsilon}$,
\begin{equation}
\sigma=\frac{16\pi e^2m}{3\hbar^2}M_1,
\end{equation}
where $M_n=\int^{0}_\infty l(\epsilon)\epsilon^n\frac{\partial f_0}
{\partial \epsilon}$; $l(\epsilon)$ is a length of free motion of the
electrons.


Similarly it is possible to show that the nonequilibrium distribution
function for the electron gas in the stationary gravitational field
in the presence of magnetic field
leads to the general relativistic Ohm's law

\begin{eqnarray}
\label{grohm}
F_{\alpha\beta}u^{\beta}=\frac{1}{\sigma} j _{\alpha}+
R_H(F_{\nu\alpha}+u_\alpha u^\sigma F_{\nu\sigma}) j^\nu
+\Xi^{-1/2}\stackrel{\perp}{\nabla}_\alpha\tilde\zeta_e\nonumber\\ -
\beta\Xi^{-1/2}\stackrel{\perp}{\nabla}_\alpha\tilde
T-R_{gg} j^\beta A_{\alpha\beta} \ ,
\end{eqnarray}
where $\beta$ is the thermoelectric power,
$R_H$ is the Hall constant,
$R_{gg}=2mc/ne^2$ is galvanogoromagnetic coefficient,
$u_\alpha$ is the four-velocity of the conductor as whole,
$A_{\beta\alpha}=u_{[\alpha,\beta]}+u_{[\beta} w_{\alpha ]}$ is the
relativistic rate of rotation,
$\stackrel{\perp}{\nabla}_\alpha$ is the transversal part of the covariant
derivative,
[...] denotes antisymmetrization.

One can estimate that in the weak field approximation the
following relation is valid$^{(13)}$
\begin{equation}
\stackrel{\perp}{\nabla}_\alpha\tilde\zeta_e
=Aw_\alpha=a-\gamma M_ac^2/e\ ,
\end{equation}
where $a=mc^2/e$, $\gamma$ is parameter of order of $1$,
$M_a$ is atomic mass$^{(13)}$.

\section{GYROSCOPIC AND GRAVITOMAGNETIC EFFECTS IN
CURRENT CARRYING CONDUCTORS}

     The charge distribution and total electric  charge  quantity
in the rotating conductors carrying current has been considered
in$^{(6)}$. According to this paper a `gyroscopic' effect
is expected to exist for conductor carrying a direct current. That is opposite
electric charges to be  produced  at
the different parts of  a  (constant)  current  carrying  multiturn
rotating solenoid provided  that  coil  has  circular
rings of wire with opposite  directions.  If  the
current aligned along the linear  velocity  of  rotation  coincide  then  the
conductor takes positive charge and vice versa. On the  base
of the predicted `gyroscopic' effect it has been proposed$^{(6)}$
to  detect an
alternating electric current of `charge  exchange' through series-connected
current  carrying  solenoids  with   opposite   windings   during
the oscillatory motion around their own axis.

Probably
the `gyroscopic' effect in the sense of the paper$^{(6)}$ probably
could not exist  due to the following
two reasons: first, the appearance of  the  surface  charge  density
$\sigma$
besides  the  space  one $\rho_0$ has  been  ignored in$^{(6)}$,
second,  the expressions being valid only in the flat space-time, for
example, transmission formulae  from the  field  tensors  to  observable
quantities or the Ohm's law in standard form have been groundlessly used
 for the calculations in the rotating frame of reference.

     Our aim is to examine the theoretical proposal on measuring
`gyroscopic' effect in series-connected two solenoids carrying azimuthal
   current in opposite directions$^{(6)}$   in  the
rotating frame of reference

\begin{eqnarray}
\label{rotating}
ds^2=-(c^2-\Omega^2 r^2)dt^2+2\Omega r^2d\varphi dt+dr^2+
r^2d\varphi^2+dz^2 ,
\end{eqnarray}
where $\Omega$ is angular velocity of rotation, $x^1=r, x^2=\varphi,
x^3=z$ are the cylindrical coordinates.

     The general relativistic expression for  the  proper  charge
density $\rho _0$  inside conductors$^{(7)}$
\begin{eqnarray}
\label{chargedensity}
&&\rho_0 = \frac{\epsilon\mu R_H}{c} j^2+\frac{1}{4\pi}
    \{(\frac{\epsilon}
        {\sigma} j^\alpha)_{;\alpha}+[\epsilon^2\mu
        R_H(\frac{1}{\sigma} j^2+
        {\Xi^{-1/2}} j^\nu\stackrel{\perp}{\nabla}_\nu
    (\Xi^{1/2}\zeta_e))u^\alpha ]_{;\alpha}
\nonumber\\
&&-\epsilon R_{gg}A_{\alpha\beta}
w^\alpha j^\beta+
g^{\alpha\beta}(\epsilon R_{gg} j^\nu
A_{\alpha\nu})_{;\beta}-
\frac{\epsilon}{\sigma}w^\alpha j_\alpha -
\epsilon w^\alpha{\Xi^{-1/2}}
\stackrel{\perp}{\nabla}_\alpha(\Xi^{1/2}\zeta_e)\nonumber\\
&&+ g^{\alpha\beta}(\epsilon {\Xi^{-1/2}}
\stackrel{\perp}{\nabla}_\alpha(\Xi^{1/2}\zeta_e))_{;\beta}
+H^{\alpha\beta}[A_{\beta\alpha}+\epsilon\mu R_Hw_\alpha j_\beta
   +(\epsilon\mu R_H j_\alpha)_{,\beta}]\}
\end{eqnarray}
has been derived from the Maxwell equations (\ref{maxwell})
assuming that the constitutive relations between fields and
inductions have linear character as well as using the generalized
Ohm law (\ref{grohm}). Here $\epsilon$ and $\mu$ are the
parameters for the conductor.

     In the general case integral$^{(13)}$

\begin{eqnarray}
Q = -\frac{1}{c}\int (J^\alpha)_*dS_\alpha = -\frac{1}{c}\int J^\alpha
r_\alpha dV = \int \rho _v dV,
\end{eqnarray}
on any hypersurface $dS^{\alpha\beta\gamma}$ is equal to the sum of electric
charges whose world-lines intersect this hypersurface. Here element

\begin{equation}
*dS_\sigma =\frac{1}{3!}dS^{\alpha\beta\gamma}e_{\alpha\beta\gamma\sigma}=
\frac{e_\sigma -(ek)k_\sigma}{\sqrt{1+(ek)^2}}dV= r_\sigma dV
\end{equation}
is dual to the hypersurface
built on the triple of vectors $\{{\mathbf k}, {\mathbf m}, {\mathbf n}\}$

\begin{equation}
m_\alpha =\frac{e_{\lambda\alpha\mu\nu}e^\lambda n^\mu k^\nu}
{\sqrt{1+(ek)^2}}, n_\alpha =\frac{e_{\lambda\alpha\mu\nu}
e^\lambda k^\mu m^\nu}{\sqrt{1+(ek)^2}}, k^\alpha=-(ek)e^\alpha+
\sqrt{1+(ek)^2}e^{\mu\alpha\rho\nu}e_\mu m_\rho n_\nu , \nonumber
\end{equation}
$\rho _v$  is the charge density per unit volume, time-like  unit  vector
$r^\alpha$  can be represented in the form

\begin{equation}
r_\alpha =e_\alpha \sqrt{1-v^2/c^2}-v_\alpha /c,
\end{equation}
$v_\alpha$ is the velocity of observer with respect to object
$\{{\mathbf k},{\mathbf m},{\mathbf n}\}$, $k_\alpha =v_\alpha /v$ that is
$(ek)=\frac{v/c}{\sqrt{1-v^2/c^2}}$, $v=\sqrt{v_\alpha v^\alpha}.$

     Assume an infinitely long hollow cylindrical ($r_1$  and
$r_2$ are  radii  of  interior  and  exterior  surfaces) conductor
carrying current in azimuthal direction ($I$ is the current per
the unit length of the cylinder) is in the rotating frame of
reference ~(\ref{rotating}); see Figure 1.

\begin{figure}
\begin{center} \label{fig1}
\includegraphics[height=60mm, width=86mm]{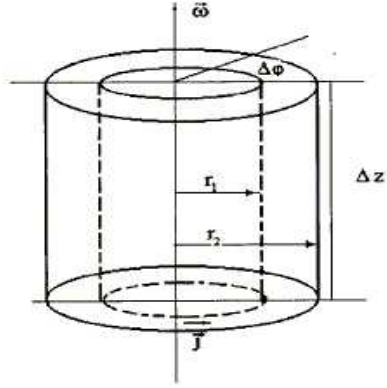}
\end{center}
\caption{A piece of infinite
hollow  cylinder  carrying current in azimuthal direction in the
rotating frame of reference.}
\end{figure}

     Due to the  cylindrical  symmetry  of  the  problem  Maxwell
equations (2) have the following nonvanishing solutions: \\
in the cavity of the conductor $(0<r<r_1)$

\begin{eqnarray}
B_z= \frac{4\pi I}{\sqrt{c^2-\Omega^2r^2}},\quad
E_r=\frac{\Omega r}{c} \frac{4\pi I}{\sqrt{c^2-\Omega^2r^2}};\nonumber
\end{eqnarray}
inside conductor $(r_1\le r\le r_2)$

\begin{eqnarray}
\label{h_z}
H_z=  \frac{4 \pi I}{\sqrt{c^2-\Omega^2r^2}}\frac{\ln (r_2/r)}
{\ln (r_2/r_1)}&,&\\
\label{e_r}
E_r=A\frac{\Omega^2 r}{c^2-\Omega^2 r^2}
-\frac{R_{gg}I\Omega}{r(1-\Omega^2r^2)^2\ln (r_2/r_1)}&,&\\
\label{zaryad}
\rho_0= \frac{2\Omega c I}{(c^2 -\Omega^2r^2)^{3/2}} \frac{\ln (r/r_2)}
{\ln (r_2/r_1)}+\frac{1}{4\pi}\Biggl\{ \frac{\epsilon A \Omega^2(2c^2+
\Omega^2r^2)}{(c^2-\Omega^2 r^2)^2}+\frac{\Omega^2 r}{c^2-\Omega^2 r^2}
\frac{\partial(\epsilon A)}{\partial r}\Biggr\} &&\nonumber\\
-\frac{1}{4\pi}\Biggl\{\frac{I\Omega}{r(1-\Omega^2r^2)^2\ln (r_2/r_1)}
\frac{\partial (R_{gg}\epsilon)}{\partial r} +
\frac{5R_{gg}\epsilon I\Omega^3}{(1-\Omega^2r^2)^3\ln (r_2/r_1)}\Biggr\}&.&
\end{eqnarray}
Thus according to the generalized Ohm's law~(\ref{grohm}) and the
continuity of the electric current at the boundaries
$r=r_1$ and $r=r_2$, the internal radial electric field
~(\ref{e_r}) has two contributions: one is due to the absolute acceleration
being proportional to $\Omega^2$ and the second one results from the rotational
effect on azimuthal current and is linear in angular velocity of rotation.
The nonvanishing space charge~(\ref{zaryad}) generated inside the solenoid
is due to three different relativistic reasons: the first term on the left
hand side of~(\ref{zaryad}) is produced by the general relativistic effect of
charge redistribution$^{(7)}$ in the presence of magnetic field
~(\ref{h_z}), the second one is produced by the inner electric field being
proportional to absolute acceleration, and the last one is arising from the
interplay between azimuthal electric current and the rotational field.

     Densities of surface charges induced at the surfaces $r_1$ and
$r_2$ of the conductor

\begin{eqnarray}
  \sigma_{|r=r_1}&=&-\frac{\Omega r_1I}{c\sqrt{c^2-\Omega^2r_1^2}}+
\frac{\epsilon A}{4\pi}\frac{\Omega^2 r_1}{c^2-\Omega^2 r_1^2}
-\frac{R_{gg}\epsilon I\Omega}{4\pi r_1(1-\Omega^2r^2)^2\ln (r_2/r_1)}
,\\
\sigma_{|r=r_2}&=&-\frac{\epsilon A}{4\pi}\frac{\Omega^2 r_2}{c^2-\Omega^2
r_2^2}-\frac{R_{gg}\epsilon I\Omega}{4\pi r_1(1-\Omega^2r^2)^2\ln (r_2/r_1)}
\end{eqnarray}
have been found from the boundary conditions for  the vectors  of
electromagnetic field.

     The total quantity of space charge

\begin{eqnarray}
Q_\rho =\frac{\epsilon A}{4\pi}\left[\frac{\Omega^2r^2_2/c^2}{(1-
\Omega^2r^2_2/c^2)^{3/2}}-\frac{\Omega^2r^2_1/c^2}{(1-\Omega^2r^2_1/c^2)
^{3/2}}\right]
\Delta z\Delta\varphi +\frac{I\Omega r_1^2}{c^2-\Omega ^2r_1^2}
\Delta z\Delta\varphi \nonumber\\
-\frac{R_{gg}\epsilon I\Omega}{4\pi\ln (r_2/r_1)}\left[
\frac{1}{(1-\Omega^2r_2^2)^{5/2}}-\frac{1}{(1-\Omega^2r_1^2)^{5/2}}\right]
\Delta z\Delta\varphi
\end{eqnarray}
induced inside an arbitrary piece of the solenoid limited by two
2-surfaces $\varphi = const.$ ($\varphi =\varphi _1$ and $\varphi =
\varphi _2$, $\Delta\varphi =\varphi _2-\varphi _1$) and two 2-surfaces
$z= const.$ ($z=z_1$ and $z=z_2$ , $\Delta z=z_2 -z_1$) is completely
compensated by the total quantity of surface charges

\begin{eqnarray}
Q_\sigma =\frac{\epsilon A}{4\pi}(\frac{\Omega^2r^2_1/c^2}{(1-\Omega^2r^2_1/
c^2)^{3/2}}-\frac{\Omega^2r^2_2/c^2}{(1-\Omega^2r^2_2/c^2)^{3/2}})
\Delta z\Delta\varphi -\frac{I\Omega r_1^2}{c^2-\Omega ^2r_1^2}
\Delta z\Delta\varphi \nonumber\\
-\frac{R_{gg}\epsilon I\Omega}{4\pi\ln (r_2/r_1)}\left[
\frac{1}{(1-\Omega^2r_1^2)^{5/2}}-\frac{1}{(1-\Omega^2r_2^2)^{5/2}}\right]
\Delta z\Delta\varphi
\end{eqnarray}
generated at the pieces of surfaces $r=r_1$ and $r=r_2$.

     Thus in spite of violation of the local electric  neutrality
net sum of the space and surface charges of conductor
$Q = Q_\rho + Q_\sigma $ is identically
equal to zero and the `charge exchange' current due to the angular vibration
of two series-connected coil conductors carrying azimuthal current in
opposite directions$^{(6)}$ does not exist. The suggested proposal$^{(6)}$
on measuring this 'charge exchange' current has to give the null result.

However, we may note that in our recent papers$^{(15)}$
we theoretically predicted the galvanogravitomagnetic and
galvanogyroscopic voltages produced by the effect of gravitomagnetic and
rotational fields on radial conduction current. Moreover, according
to our theoretical results$^{(16)}$, the experiment of Vasil'ev$^{(16)}$
on measurement of vertical magnetic field around rotating
hollow neutral cylinder carrying radial thermoelectric current confirms the
existence of gyroscopic effects on electric current.

It is naturally to ask the question about a possibility to measure
gyroscopic effect in the rotating solenoid carrying azimuthal current.
Since, even in the linear in $\Omega$ approximation the inner radial
electric field $E_r=-{R_{gg}I\Omega}/{r\ln (r_2/r_1)}$ is
produced inside the solenoid, one can suggest the following non-null
proposal on measuring gyroscopic voltage in radial direction. Suppose
that  the superconducting wire is connected to the inner and outer sides
of the solenoid in the way when it forms superconducting loop with SNS
junction where the voltage is produced. Then the nonvanishing potential
difference $V_{r}$ would lead to a time
varying magnetic flux through the loop. The change in magnetic flux
$\Phi_b$ inside the circuit during the time interval $[0,t]$ is $^{(18)}$

\begin{eqnarray}
\Delta\Phi_b =\Delta n\Phi_0+ {c}\int_{0}^{t}V_{r}dt\ ,
\label{eq:flux}
\end{eqnarray}
where $\Phi_0=\pi\hbar c/e=2\times 10^{-7}Gauss\cdot
cm^2$ is quantum of the magnetic flux.
 As long as $\Delta\Phi_b<\Phi_0$,
$n$ will remain constant and $\Delta\Phi_b$ will increase linearly with
time until $\Delta\Phi_b=\Phi_0$, then the order of the step $n$ will
change as flux quantum enters the loop.
Thus this particular loop is sensitive to the $V_{r}$ and in this
connection to the angular velocity of rotation.

If the conductor is placed on a platform that is nonrotating relative to
the distant stars as determined by telescopes, then, in general, it
would rotate with respect to the local inertial frames if a nearby
rotating body as the Earth was present because of the general-relativistic
Lense-Thirring field which is a consequence of gravitational mass
currents$^{(19, 20)}$. If the Earth is approximately spherically symmetric,
the Lense-Thirring angular velocity of the local inertial frames relative
to the distant stars at position, ${\mathbf r}$, from the centre of the
Earth is

\begin{equation}
{\mathbf\Omega_{LT}}=\frac{4GM_\oplus R^2_\oplus}{5c^2}\left[
-\frac{{\mathbf\Omega}_\oplus}{r^3}+\frac{{(\mathbf\Omega}_\oplus
{\mathbf r)r}}{r^5} \right],
\end{equation}
where $M_\oplus$, $R_\oplus$, and ${\mathbf\Omega}_\oplus$ are the mass,
radius and angular velocity of the Earth, respectively, and $G$ is the
gravitational constant.  As a rule, the platform on the Earth is rotating
with respect to the distant stars, but one can expect that
the other effects of rotation may be eliminated with the additional methods.

Assume that a piece of a conductor is at rest on the platform and connected
to the source of alternating current in azimuthal direction $i_\varphi =
i_\varphi (0) \exp{(i\omega t)}$ with the frequency $\omega$.
Hence, the alternating radial galvanogyroscopic current
\begin{equation}
i_r=i_\varphi (0) \exp{(i\omega t)}\frac{R_{gg}\sigma\Omega_{LT}}
{c}
\end{equation}
will be produced across the conductor according to the general-relativistic
Ohm's law~(\ref{grohm}). If the amplitude of the supplied azimuthal
current is about $10^3 A$,
$R_{gg}=10^{-22}s$, $\sigma =10^{17}c^{-1}$ the developed azimuthal
current has amplitude around $10^{-21}A$.
Such currents can be measured, however, in the present case there are
serious environmental problems which could reduce the feasibility of the
experiment.

\section{CONCLUSION}

We conclude:

 \noindent 1. The solution of the general-relativistic
Boltzmann's kinetic equations for the nonequilibrium distribution
function for electron gas leads to the Ohm's law for conduction
current being valid in the fields of gravity and inertia. The
latter has pure general-relativistic contributions arising from
the absolute acceleration and relativistic rate of rotation.

\noindent
2. `Gyroscopic' effect in the sense of the paper$^{(6)}$ does not
exist since rotationally induced space charges in the solenoid with
electric current are totally compensated by the surface charges
generated at the
inner $r=r_1$ and the outer $r=r_2$ boundaries of the solenoid.

\noindent
3. Experiment is proposed to measure radial voltage through the rotating
solenoid carrying azimuthal electric current.

\noindent
4. Finally, the nonstationary situation, when alternate azimuthal current
produces radial current across the conductor due to the effect of the
Earth's Lense-Thirring field has been considered.

\section*{ACKNOWLEDGEMENT}

BA and ME thank the IUCAA for warm hospitality during their stay
in
 Pune, AS-ICTP and TWAS for the travel support. This research is
also supported in part by the UzFFR (project 01-06) and projects
F.2.1.09, F2.2.06 and A13-226 of the UzCST. BA acknowledges the
partial financial support from NATO through the reintegration
grant EAP.RIG.981259.

\newpage
{\bf REFERENCES}\\
\begin{enumerate}
\item
J. Ehlers, ``General relativity and kinetic theory", in
{\it General Relativity and Cosmology},
ed. R.K. Sachs (Academic Press, 1971) p.1.
\item
A. Georgiou, ``Relativistic transport equations for plasmas",
{\it Class. Quantum Grav.} {\bf 12}, 1491-1501 (1995).
\item
G. Brodin, and M. Marklund, ``Parametric excitation of plasma waves by
gravitational radiation",
{\it Phys. Rev. Lett.} {\bf 82}, 3012-3015 (1999).
\item
K. Els\"asser, and S. Popel, ``Plasma equations in general relativity",
{\it Phys. Plasmas} {\bf 4}, 2348-2356 (1997).
\item
D.C. Kelly, ``Electrical and thermal conductivities of a relativistic
degenerate plasma", {\it Astrophys. J.} {\bf 179}, 599-606 (1973).
\item
V.F. Fateev, ``Gyroscopic effect in coil conductors carrying electric
current", {\it Sov. Tech. Phys. Lett.} {\bf 15}, 72-75 (1989).
\item
C.W. Misner, K.S. Thorne and J.A. Wheeler, {\it Gravitation} (San Francisco,
W.H. Freeman and Company, 1973).
\item
G. Montani, R. Ruffini and R.M. Zalaletdinov, ``Gravitating macroscopic
media in general relativity and macroscopic gravity",
{\it Nuovo Cim. B} {\bf 115}, 1343-1354 (2000).
\item
A.N. Kaufman, ``Maxwell equations in nonuniformly moving media",
{\it Ann. Phys.} {\bf 18}, 264-273 (1962).
\item
N.A. Chernikov, ``The relativistic gas in the gravitational field",
{\it Acta Phys. Polon.} {\bf 23}, 629-645 (1963);
  ``Microscopic foundation of relativistic hydrodynamics",
{\it Ibid.} {\bf 27}, 465-489 (1964).
\item
S.R. de Groot, W.A. van Leuween and Ch.G. van Weert,
{\it Relativistic Kinetic Theory} (Amsterdam, North Holland, 1980).
\item
      T.W. Darling, F. Rossi, G.I. Opat, and
G.F.Moorhead
``The fall of charged Particles
under gravity: A study of expremental
problems", {\it Rev. Mod. Phys.} {\bf 64}, 237-257 (1992).
\item
       J. Anandan, ``Relativistic thermoelectromagnetic gravitational
    effects in normal conductors and
superconductors" {\it Phys. Lett. A.} {\bf 105}, 280-284 (1984).
\item
B.J. Ahmedov and L.Ya. Arifov, ``Principles for detecting charge
redistribution produced by fields of gravity and inertia inside conductors"
{\it Gen. Rel. Grav.} {\bf 26}, 1187-1195 (1994).
\item
B.J. Ahmedov, ``General relativistic galvano-gravitomagnetic effect
in current carrying conductors",
{\it Phys. Lett. A.} {\bf 256}, 9-14 (1999).
\item
B.J. Ahmedov, ``General relativistic Ohm's law and Coriolis force
effects in rotating conductors", {\it Gravit. Cosmology} {\bf 4},
139-141 (1998).
\item
B.V. Vasil'ev, ``Thermogyromagnetic effects",
{\it Russian Physics JETP Letters} {\bf 60}, 47-50 (1994).
\item
A.K. Jain, J.E. Lukens and J.S. Tsai, ``Test for relativistic gravitational
         effects on charged particles", {\it Phys. Rev. Lett.}
{\bf 58}, 1165-1168 (1987).
\item
J. Lense, H. Thirring, ``Uber den Einfluss der Eigenrotation der Zentralkorper
auf die Bewegung der Planeten und Monde nach der Einsteinschen
Gravitationstheorie",  {\it  Phys. Z.} {\bf 19}, 156-163 (1918).
\item B. Mashhoon, F.W. Hehl, D.S. Theiss, ``On the gravitational effects of
rotating masses: The Thirring-Lense papers", {\it Gen. Rel. Grav.}
{\bf 16}, 711-750 (1984).
\end{enumerate}

\end{document}